\documentclass[conference]{IEEEtran}
\IEEEoverridecommandlockouts
\usepackage{nicematrix}
\usepackage{arydshln} % For dashed lines
\usepackage{mathtools}

\usepackage{booktabs}
\usepackage{array}

\usepackage{tabularray}
\usepackage{arydshln}
\usepackage{arydshln,leftidx,mathtools}

\usepackage{siunitx}
\usepackage{amsmath,amssymb,amsfonts}
\usepackage{algorithmic}
\usepackage{graphicx}
\usepackage{textcomp}
\usepackage{xcolor}
\usepackage{amsmath}
\usepackage{amssymb}
\usepackage{mathtools}
\usepackage{amsthm}
\usepackage{amsfonts}
\usepackage{stackengine}

\usepackage{float}
\usepackage{amsmath,amsthm,amsfonts,amscd,amssymb, arydshln}
\usepackage[mathscr]{eucal}
\usepackage{graphics,graphicx,multicol}
\usepackage{epsfig}
\usepackage{diagbox}
\usepackage{subfigure}
\usepackage{stfloats}
\usepackage{latexsym}
\usepackage{amsfonts}
\usepackage{cite}
\usepackage{algorithm,algorithmic}
\usepackage{color}
\usepackage{xcolor}
\usepackage{multirow}
\usepackage{booktabs}

\usepackage[utf8]{inputenc} % usually not needed (loaded by default)
\usepackage[T1]{fontenc}

\DeclareMathOperator*{\argmin}{arg\,min}
\title{}
\author{}
\date{}

\def\BibTeX{{\rm B\kern-.05em{\sc i\kern-.025em b}\kern-.08em
    T\kern-.1667em\lower.7ex\hbox{E}\kern-.125emX}}
\begin{document}

\title{Sensing Management for Pilot-Free Predictive Beamforming in Cell-Free Massive MIMO Systems
% Predictive Resource Allocation for ISAC: 
\thanks{This work was supported by the SUCCESS project funded by the Swedish Foundation for Strategic Research.}
}
\author{
\IEEEauthorblockN{Eren Berk Kama, Murat Babek Salman, Isaac Skog, and Emil Björnson}\\
\IEEEauthorblockA{\textit{Division of Communication Systems}, \textit{KTH Royal Institute of Technology}, Stockholm, Sweden \\
Emails: \{ebkama, mbsalman, skog, emilbjo\}@kth.se}
}
% \author{\IEEEauthorblockN{Eren Berk Kama}
% \IEEEauthorblockA{\textit{Division of Communication Systems} \\
% \textit{KTH Royal Institute of Technology}\\
% Stockholm, Sweden \\
% ebkama@kth.se}
% \and
% \IEEEauthorblockN{Murat Babek Salman}
% \IEEEauthorblockA{\textit{Department of Intell. Commun. Engineering} \\
% \textit{KTH Royal Institute of Technology}\\
% Stockholm, Sweden \\
% mbsalman@kth.se}
% \and
% \IEEEauthorblockN{Isaac Skog}
% \IEEEauthorblockA{\textit{Department of Intell. Commun. Engineering} \\
% \textit{KTH Royal Institute of Technology}\\
% Stockholm, Sweden \\
% skog@kth.se}
% \and
% \IEEEauthorblockN{Emil Björnson}
% \IEEEauthorblockA{\textit{Division of Communication Systems} \\
% \textit{KTH Royal Institute of Technology}\\
% Stockholm, Sweden \\
% emilbjo@kth.se}
% }
\maketitle
\begin{abstract}
% \textcolor{red}{I will further paraphrase the abstract}
% This paper presents a novel integrated sensing and communication (ISAC) framework for cell-free massive MIMO systems that operate on a state-based approach. States are defined based on the communication request of the user and the angle of it is tracked when there is no communication need. We consider predictive precoding and propose an approach to eliminate the need for continuous channel estimation, substantially reducing the communication overhead. We develop an extended Kalman filter (EKF) based approach to track the target angle. In addition,  resource allocation strategies including predictive sensing time selection and receive access point (AP) selection are proposed to enhance tracking accuracy and improve the overall signal-to-noise ratio (SNR) during downlink communication. Numerical results validate the efficacy of the proposed ISAC framework, demonstrating the possibility of overhead-free communication with predictive beamforming.
% This paper introduces a novel integrated sensing and communication (ISAC) framework for cell-free massive MIMO systems. 
This paper introduces a sensing management method for integrated sensing and communications (ISAC) in cell-free massive multiple-input multiple-output (MIMO) systems. Conventional communication systems employ channel estimation procedures that impose significant overhead during data transmission, consuming resources that could otherwise be utilized for data. To address this challenge, we propose a state-based approach that leverages sensing capabilities to track the user when there is no communication request. Upon receiving a communication request, predictive beamforming is employed based on the tracked user position, thereby reducing the need for channel estimation. Our framework incorporates an extended Kalman filter (EKF) based tracking algorithm with adaptive sensing management to perform sensing operations only when necessary to maintain high tracking accuracy. The simulation results demonstrate that our proposed sensing management approach provides uniform downlink communication rates that are higher than with existing methods by achieving overhead-free predictive beamforming.
\end{abstract}
\section{Introduction}
% \textcolor{red}{I am still writing the Introduction section, please skip it for now.}

In recent years, the combination of wireless communications and sensing has attracted considerable attention \cite{10217169}. In the current infrastructure, radar and communication systems are operated independently, each with its dedicated hardware and spectrum allocation \cite{9737357}. However, integrated sensing and communications (ISAC) seeks to unify these functionalities to achieve multifunctional wireless systems with joint resource allocation and enhanced performance for both services.

% There are several works considering predictive beamforming in joint sensing communication literature mainly assuming a massive MIMO setting. Bayesian methods to predict vehicle motion from radar-communication signal echoes, reducing signaling overhead in vehicular networks are given in \cite{9246715}.  Sensing is used to track and predict vehicle movement, to decrease overhead in vehicle-to-infrastructure communication in a massive MIMO setting in \cite{9171304}. EKF-based predictive beamforming for tracking a drone in 3D is proposed in \cite{10214383}.  Deep learning is applied to predict beamforming matrices, enhancing communication rate while lowering channel tracking overhead in \cite{9791349}. \cite{9492131} Introduces a deep learning-based approach to predictive beamforming that reduces training data dependency. \cite{akçalı2025predictivebeamformingdistributedmimo} proposes predictive beamforming strategies for distributed MIMO systems, which enhance throughput while reducing overhead. Previous works consider transmitting sensing and communication signals at the same time.

Channel estimation is employed in most communication systems, which consume resources that could otherwise be allocated to data transmission. Prior work has addressed this issue by employing user tracking and predictive beamforming in propagation scenarios with smooth movements. For instance, a Bayesian tracking method was applied to predict vehicle motion from radar echoes, thereby reducing signaling overhead in vehicular networks \cite{9246715}. Sensing-based vehicle tracking has also been utilized to decrease overhead in vehicle-to-infrastructure communication in a massive multiple-input multiple-output (MIMO) setting \cite{9171304}. Moreover, extended Kalman filter (EKF)-based predictive beamforming for three-dimensional drone tracking was presented in \cite{10214383}. Deep learning techniques also have been proposed to predict beamforming matrices, enhancing communication rates and reducing channel estimation overhead \cite{9791349,9492131}. Predictive beamforming strategies for distributed MIMO systems are presented in \cite{akçalı2025predictivebeamformingdistributedmimo}. 
% Effective ISAC necessitates a waveform supporting both sensing and communication without performance degradation, posing a significant design challenge. Recent investigations focus on integrating MIMO-OFDM into ISAC to exploit its communication advantages for dual-purpose systems. 

%%%%%%%%%% FIGURES
% A model of the considered system can be seen in Fig. \ref{fig:APs and car system}.

\begin{figure}[t!]
\centering
        \includegraphics[scale=.1,width=8.7cm]{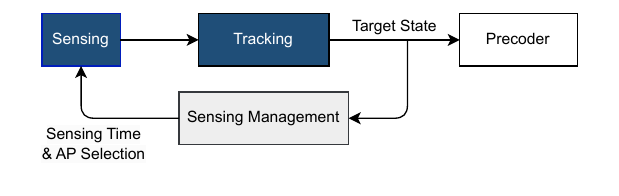}\vspace{-3mm}
    \caption{Conceptual diagram of the proposed predictive beamforming method. Sensing and tracking provide an estimate of the user position information when needed and the precoder is formed on demand. }
    \label{fig:systemmodel}\vspace{-3mm}
\end{figure}

Existing solutions assume simultaneous transmission of sensing and communication signals, with a predominant focus on massive MIMO architectures that yield non-uniform spectral efficiency and full-buffer data transmission. This paper explores predictive beamforming using user tracking in cell-free massive MIMO systems, where user communication is initiated on demand rather than continuously maintained. This creates a state‐based communication framework, where data is transmitted in bursts, eliminating the need for radar and communication signals to be sent simultaneously. By leveraging position information obtained through user tracking during dedicated sensing periods, the proposed ISAC framework removes the need for channel estimation to perform beamforming, thereby significantly reducing the associated overhead when users request access. Moreover, by using the ability of the user tracking filter to predict the user position and assess the uncertainty of the predictions, a novel sensing management method is proposed. This method controls the sensing so that sensing signals are only transmitted when needed to maintain high tracking precision, thereby reducing resources allocated for sensing.
% In contrast, our contributions leverage radar tracking within a cell-free massive MIMO framework that supports state‐based, burst-mode communication. By exploiting angle information from dedicated sensing periods, the proposed scheme obviates the need for channel estimation, thereby significantly reducing signaling overhead.}

 % A scheme of the considered system can be seen in Fig. \ref{fig:systemmodel}. By using a cell-free massive MIMO setup, we aim to enhance the delay and Doppler shift estimation performance to obtain accurate range and velocity information of the target.
Fig.~\ref{fig:systemmodel} illustrates the considered system setup. Utilizing a cell-free massive MIMO architecture, our objective is to keep track of the target's location (e.g., range and velocity) by optimized delay and Doppler shift estimation. We perform delay and Doppler shift estimation using 2D FFT and OFDM waveforms. We derive the EKF state equations for the cell-free massive MIMO model to track the user position and velocity, which in turn facilitates accurate angle prediction for precoding. Moreover, based on the predicted angle estimation error, we propose a predictive selection of receive access points (APs), whereby the observed signal varies with each receiver (Rx) AP set. Finally, we determine how frequently sensing needs to be done to maintain sufficient estimation accuracy. Numerical results demonstrate that sensing need not be conducted frequently, and communication in bursts permits the implementation of predictive beamforming, which effectively eliminates the channel estimation overhead.

\begin{figure}[t!]
\centering
        \includegraphics[scale=.01,width=8.7cm]{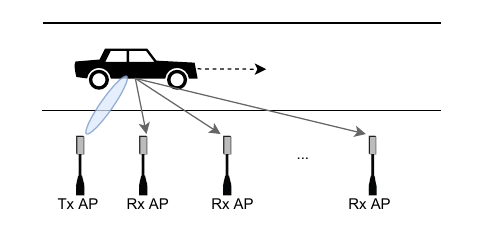}\vspace{-3mm}
    \caption{The considered system where the first AP is the transmitter (Tx) AP and a set of Rx APs are selected for sensing reception.}
    \label{fig:APs and car system}\vspace{-3mm}
\end{figure}

% and that predictive beamforming effectively eliminates channel estimation overhead as communication is in bursts.
% Numerical results indicate that sensing can be performed infrequently, and burst-mode communication permits the implementation of predictive beamforming, which effectively eliminates the channel estimation overhead.

% We do delay doppler estimation with 2D FFT,
% We use OFDM waveform, 
% We aim to improve the estimation quality
% We derive the EKF state equations for the cell-free massive MIMO model
% We track the position and velocity to find angle. 
% We predict the angle and form precoder
% We find the predicted angle estimation error and decide which APs should be Rx APs- predictive allocation of Rx APs.  - which changes - all the tracking and estimation through the received signal - the observed signal changes at every Rx AP set
% Then we decide how frequently sensing should be performed.

% Numerical results show that we don't need to do sensing too frequently. Also, as the communication is in bursts, we can use predictive beamforming, which eliminates channel estimation overhead.

\section{Communication and Sensing System Model}

We consider a cell-free massive MIMO ISAC system.
There are $L_{\rm{T}}$ APs, each with $N$ antennas, and single-antenna users. The APs are controlled by a central processing unit (CPU) and phase-synchronized to enable joint transmission and reception.

User traffic is inherently bursty in practice, leading to frequent transitions between active and idle states, even during continuous usage of user applications. A conventional network typically loses track of the user's location when it enters the idle mode. To overcome this, we leverage sensing such that when data is not transmitted, APs use sensing signals to estimate and track the user position, enabling seamless service continuity. 
% When there is no communication request and no need for angle of departure estimation, the user is in the idle state. This state can be used to sense or serve another user if multiple users are considered.  Downlink communication and sensing are done during separate time epochs, which we denote with $\kappa$. %We use sensing to track the range, velocity, and angle of the target.
For brevity, we focus on the communication and tracking of a single user in this paper.

%Sensing signals are transmitted when there is a need to enhance the accuracy of the user position estimate. 

We use OFDM as the common waveform for both communication and radar signals throughout the paper. 
% We use the time samples and the symbols instead of the fast-time and slow-time throughout the paper.
The OFDM signal at symbol time $b$ can be written as
\begin{equation} \label{eq:OFDM pulse}
\varsigma_{b}[m] =  \frac{1}{\sqrt{N_c}} \sum_{a=0}^{N_c-1} \gamma_{a,b} e^{j2\pi a \frac{m}{N_c}},
\end{equation}
where $N_c$ is the number of subcarriers  and $a$ and $b$ are the subcarrier and symbol indices, respectively. In a coherence block consisting of $N_{s}$ OFDM symbols, the time samples and symbols have the range $m=0,\ldots,N_{c}-1$ and $b=0,\ldots,N_{s}-1$, respectively. We denote the transmitted data symbols and the code for radar as $\gamma_{a,b}$. To mitigate inter-symbol interference, the signal $\varsigma_{b}[m]$ is extended by appending a cyclic prefix (CP), which is formed by taking the last $N_{cp}$ samples of $\varsigma_{b}[m]$ and placing them at the beginning of the symbol creating an $N_{c}+N_{cp}$ length signal. The CP duration $N_{cp}$ is adjusted to be larger than the maximum round-trip delay of the target. 
% The signal $\varsigma[m]$ is constructed by sequentially concatenating the $N_s$ time-domain OFDM symbols.

% We note that the transmitted data sequence $\gamma_{a,b}$ is a random variable whose realizations are known at all the APs at the time of transmission. 

\begin{figure}[t!]
\centering
        \includegraphics[scale=.5,width=8.7cm]{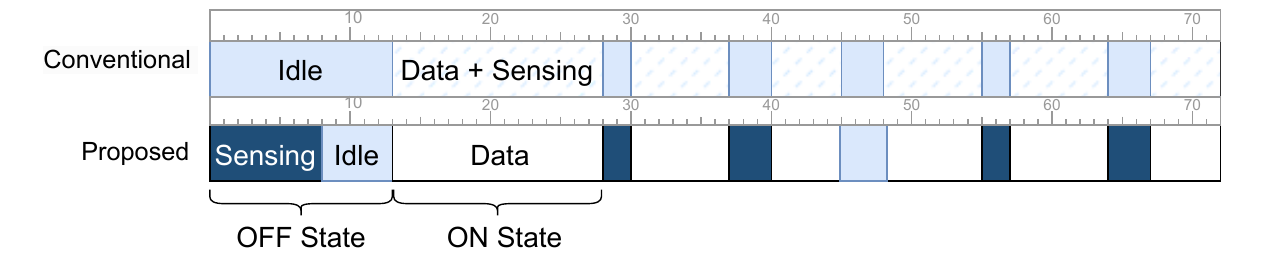}
    \caption{The conventional and proposed frame structures used in the transmission from the CPU to the UE. The sizes of blocks are chosen differently to imply the changing lengths of the states.}
    \label{fig:frame}
\end{figure}

We consider a state-based communication framework where data signals are transmitted to users upon request. The user remains in an \emph{OFF} state when it is not requesting data, during which no communication signals are transmitted to the user. Upon initiating a communication request, the user switches to the \emph{ON} state, triggering the transmission of communication signals. The state transition is controlled by the CPU. The communication signals are precoded towards the user based on the predicted user location, eliminating the need for channel estimation at the AP side. 
% Other works considering predictive beamforming assume that data streams and sensing signals are transmitted at the same time.
In the conventional frame structure, considered in previous works such as \cite{akçalı2025predictivebeamformingdistributedmimo,9171304}, the system sends both data and sensing signals when there is a communication request from the user. There is no sensing when there is no data. The proposed system sends sensing signals occasionally, but not when data is transmitted.
These structures are shown in Fig.~\ref{fig:frame} for comparison. The figure shows examples of the above-mentioned state transitions in the proposed frame structure and corresponding transitions between idle and communication states in the conventional structure. The time frame corresponds to the one of the tracking filter, which is explained in Section ~\ref{sec: EKF}. We note that, in the proposed frame structure, sensing is done until the requirement is fulfilled.

% \textcolor{purple}{What are the main contributions and novelty of this work. It is better to summarize them, and some literature review might be good to proceed. What makes it different from "Predictive Beamforming with Distributed MIMO", should be clarified.}

\subsection{Channel Model} \label{Sec: Communication Channel Model}
For the considered user, we assume there is a pure line-of-sight (LOS) channel between AP $ l $ and the user 
\begin{align}
\mathbf{h}_{l} = e^{j\varphi_{l}}\sqrt{\beta_{l}} \, \mathbf{a}(\theta_{l} ),
\end{align}
where $ \beta_{l} $ is the channel gain, $e^{j\varphi_{l}}$ is the phase-shift at the first antenna, $ \mathbf{a}(\theta_{l} ) $ is the array response vector, and $\theta_{l} $ is the angle of departure from AP $ l $ to the user in the azimuth plane. The channel gain is expressed as $\beta_{l} = \left( \lambda/ (4\pi R_{l}) \right)^2$, where $ \lambda $ is the carrier wavelength and $ R_{l} $ is the distance between AP $ l $ and the user. Assuming that horizontal uniform linear arrays (ULA) are used by all APs, the array response vector of AP $l$ is $ \mathbf{a}(\theta_{l} ) = [ 1,  e^{j \frac{2\pi}{\lambda} d \sin\theta_{l}},  \ldots,  e^{j \frac{2\pi}{\lambda} (N-1)d \sin\theta_{l}} ]^{\top}$,
where $ d $ is the antenna spacing. %and $\theta_{l}$ is the angle between the AP $l$ and the user. 

\subsection{Downlink Data Transmission}

If the system is in the \emph{ON} state, all APs serve the user in the downlink through coherent joint transmission. 
The transmitted signal $\mathbf{x}_{\bar{l}}[m]\in \mathbb{C}^{N \times 1}$ at time instance $m$ is written as
\begin{align}
    \mathbf{x}_{\bar{l},b}[m]=  \mathbf{w}_{\bar{l}} \varsigma_{b}[m], 
\end{align}
where $\mathbf{w}_{\bar{l}}\in \mathbb{C}^{N\times 1}$ is the precoder from AP $\bar{l}$ to the user and $\varsigma_{b}[m]$ is the communication signal that is the same for all APs thanks to the coherent transmission.
We assume that the data streams have unit power, i.e., $\mathbb{E}\{|\varsigma_{b}[m]|^2\}=1$. The transmitted signals should satisfy the transmit power constraint at the APs: $\mathbb{E} \left\{\Vert\mathbf{x}_{\bar{l}}[m] \Vert^{2} \right\}=  \Vert\mathbf{w}_{\bar{l}}\Vert^{2} \leq \rho_{d}$, where $\rho_d$ is the maximum AP transmit power and $\Vert \cdot \Vert$ is the $L_{2}$-norm.

For a considered coherence block where the user is in the ON state, the communication SNR can be expressed as
\begin{equation} \label{eq:SNRexpression}
\mathrm{SNR} = \frac{\left|\mathbf{h}^{\mathrm{H}}\,\mathbf{w}\right|^2}{\sigma^2_n},
\end{equation}
where $\mathbf{h}=[\mathbf{h}^{\top}_{1},\ldots,\mathbf{h}^{\top}_{L_{\rm{T}}}]^{\top}$ is the stacked channel between all APs and the user and $\sigma^2_n$ is the noise power. $\mathbf{w}=\left[\mathbf{w}_{1},\ldots,\mathbf{w}_{L_{\rm{T}}}\right]^{\top}$ is the stacked transmit array response steered to the estimates of the angle. To design effective precoders, the APs require precise knowledge of the angle, which we obtain through sensing and tracking in later sections.

\subsection{Sensing Signal Transmission}\label{subsec: Sensing Transmitted}
When the user is in the OFF state and there is a need for sensing to improve the position estimate, the system transitions into the sensing state. In this state, we assume that one AP acts as a transmitter (Tx) AP, and the set of Rx APs is decided as detailed in Section~\ref{sec: sensing management}. The remaining APs can serve in sensing or communication with other users in a multi-user scenario. Without loss of generality, we index the first AP as the transmit AP. 
% Section \ref{sec: Receive AP Selection} explains the selection of the elements of this matrix.
The transmitted radar signal $\mathbf{x}_{1}[m]\in \mathbb{C}^{N \times 1}$ at time instance $m$ can be written as
\begin{align}
    \mathbf{x}_{1,b}[m]= \mathbf{w}_{1} \varsigma_{b}[m],
\end{align}
where $\mathbf{w}_{1}\in \mathbb{C}^{N \times 1}$ is the radar precoder. 
% $\varsigma_{b}[m]$ is the transmitted radar waveform. 
We assume that the same OFDM waveform as in \eqref{eq:OFDM pulse} is transmitted for the sensing.
% as there are no communication signals in the OFF states. 
We note that the transmitted sensing signal should also satisfy the power constraint: $\mathbb{E} \left\{\Vert\mathbf{x}_{1}[m] \Vert^{2} \right\}=\Vert\mathbf{w}_{1}\Vert^{2} \leq \rho_{d}$.

When the radar signal is transmitted, the received signal $\mathbf{y}_{l,b}[m]\in \mathbb{C}^{N \times 1}$ at an Rx AP $l$, at symbol $b$ and time instant $m$ in the considered coherence block becomes
\begin{align}
&\mathbf{y}_{l,b}[m] =  \sqrt{\rho_{d}}\sqrt{\frac{\beta_{l}\beta_{1}2\pi }{\lambda^{2}}}\sigma_{l} \mathbf{a}_{}(\theta_{l})  \mathbf{a}_{}^{\top}(\theta_{1}) \mathbf{w}_{1}   \times \notag\\
&\underbrace{e^{j2\pi  b T_{\rm{sym}} \nu_{l}}\frac{1}{\sqrt{N_c}} \sum_{a=0}^{N_c-1} \gamma_{a,b} e^{j2\pi a \frac{m}{N_c}} e^{-j2\pi a \Delta_{f} \tau_{l}}}_{=\breve{\varsigma}_{b}[m](\tau_l, \nu_l)} + \mathbf{n}_{l,b}[m],
\label{eq:signal_model}
\end{align}
where $\tau_{l}$ is the propagation delay between the Tx AP and the Rx AP $l$ via the target, and $\nu_{l}$ is the Doppler shift of the target seen at receive AP $l$ and $\Delta_{f}$ is the subcarrier separation. 
%The vector $\mathbf{a}_{}(\theta_{1})\in \mathbb{C}^{N \times 1}$ is the transmit array response and $\mathbf{a}_{}(\theta_{l})\in \mathbb{C}^{N \times 1}$ is the receive array response between the receive AP $l$ and the target as defined in Section \ref{Sec: Communication Channel Model}.
The delayed and Doppler-shifted waveform is $\breve{\varsigma}_{b}[m](\tau_l, \nu_l)$.
The receiver noise is temporally and spatially white and denoted as $\mathbf{n}_{l,b}[m] \sim \mathcal{N}_{\mathbb{C}}(\mathbf{0},\sigma_n^2\mathbf{I}_{N}) \ $$ \in \mathbb{C}^{N \times 1}$, where $\mathcal{N}_{\mathbb{C}}(\cdot)$ is the circularly symmetric complex Gaussian distribution. 
% $\delta[m]$, where $\delta[m]$ is the Dirac delta function.
$\sigma_{l}$ is the radar cross section (RCS) seen from AP $l$. 
% contains the bistatic radar equation coefficient and 
The symbol duration is $T_{\rm{sym}}=\frac{1}{ \Delta_{f}}+\frac{N_{\rm{cp}}}{N_{c} \Delta_{f}}$. The scalar 
$\bar{\alpha}_{l}=\sqrt{\rho_d}\sqrt{\frac{\beta_{l}\beta_{1}2\pi }{\lambda^{2}}}\sigma_{l}   \mathbf{a}_{}^{\mathrm{H}}(\theta_{1}) \mathbf{w}_{1}$ denotes the channel gain and the inner product of the transmit array responses and the radar precoder, and will later be estimated.
% and is given by $
%     \alpha_{l,\bar{l}} = \sqrt{p^{(R)}_{t}} \sqrt{\frac{G_{tx} G_{rx} \lambda^2 }{(4\pi)^3 R_{\bar{l}}^2 R_{l}^2}}\sigma_{l}$
% where $G_{tx}$ and $G_{rx}$ are respectively the gains of the transmitting and receiving antennas, $\sigma_{l}$ is the complex radar cross section (RCS) of the target seen by AP $l$, $p^{(R)}_{t}$ is the transmitted power allocated for radar. $R_{\bar{l}}$ and $R_{l}$ are the transmitter–target and target–receiver distances, respectively.
% \textcolor{red}{\begin{align}
%     \alpha_{l} = \sqrt{\rho_d} \sqrt{\frac{G_{tx} G_{rx} \lambda^2 }{(4\pi)^3 R_{tx}^2 R_{rx}^2}}\sigma_{l}
% \end{align}}

\section{Multi-Static Sensing and its Performance}
\label{sec:sensingCRB}
In this section, we explore multi-static sensing in the described network. The Tx AP transmits a sensing signal for target localization during the sensing state, and the received signal in \eqref{eq:signal_model} is processed to obtain the estimates of the delay, Doppler shift, and angle parameters. Let 
\begin{align}
    \Phi_{l} = [\tau_{l}, \nu_{l}, \theta_{l}, \mathrm{Re}\{\bar{\alpha}_{l}\},\mathrm{Im}\{\bar{\alpha}_{l}\} ]
\end{align}
be the parameter vector. Since the noise is complex Gaussian distributed, the received signal in \eqref{eq:signal_model} can be expressed as 
\begin{align}
    \mathbf{y}_{l,b}[m] \sim \mathcal{N}_{\mathbb{C}}(\boldsymbol{\mu}_{l,b}[m;\Phi_{l}], \sigma_{n}^2 \mathbf{I}_{N} ),
\end{align}
where $\boldsymbol{\mu}_{l,b}[m;\Phi_{l}]  = \mathbb{E}\{ \mathbf{y}_{l,b}[m] \}$ is the mean of the received signal parametrized by $\Phi_{l}$. Moreover, let $\mathcal{Y}_{l}\triangleq\left\{\mathbf{y}_{l,b}[m]\right\}_{m=0,b=0}^{N_s-1,N_c-1}$ be a collection of all the time samples in a coherence block. We can write the Maximum Likelihood (ML) estimate of $\Phi_{l}$ using all the samples in $\mathcal{Y}_{l}$ as
\begin{align}
    \hat{\Phi}^{\rm{ML}}_{l}= \argmin_{\Phi_{l} \in \mathbb{R}^{5}} \sum_{m=0}^{N_c-1}\sum_{b=0}^{N_s-1}\Vert \mathbf{y}_{l,b}[m] - \boldsymbol{\mu}_{l,b}[m;\Phi_{l}] \Vert^{2},
\end{align}
 This ML estimate can be approximated by using the Discrete Fourier Transform (DFT) for OFDM waveforms \cite{9529026, 10634583, 10036975}. 
As $N_c$ and $N_s$ grow large, the ML estimate converges to
\begin{align}
    \hat{\Phi}_{l}^{\rm{ML}}\stackrel {asymp.}\sim \mathcal{N}_{\mathbb{C}}(\Phi^{*}_{l},\rm{CRB}_{\Phi_{l}}),
\end{align}
where $\Phi^{*}_{l}$ is the true value of the parameters and $\rm{CRB}_{\Phi_{l}}$ is the Cramér-Rao bound (CRB) for the parameter estimation performed at AP $l$.
% \textcolor{blue}{
% \begin{align}
%     &\text{CRB}_{\boldsymbol{\Phi}_{l}} = 
%     \begin{bmatrix}
%         \text{CRB}_{\boldsymbol{\eta}_{l}\boldsymbol{\eta}_{l}}  & \text{CRB}_{\boldsymbol{\eta}_{l}\theta_{l}} & \text{CRB}_{\boldsymbol{\eta}_{l}\alpha_{l}}\\
%         \text{CRB}_{\boldsymbol{\eta}_{l}\theta_{l}} &   \text{CRB}_{\theta_{l} \theta_{l}} & \text{CRB}_{\theta_{l}\alpha_{l}} \\ \text{CRB}_{\boldsymbol{\eta}_{l}\alpha_{l}} & \text{CRB}_{\theta_{l}\alpha_{l}} & \text{CRB}_{\alpha_{l}\alpha_{l}}
%     \end{bmatrix}.
% \end{align}}
% Fisher information matrix which can be written as 
% \begin{align}\label{eq:Fisher information matrix derivation}
%  \resizebox{\linewidth}{!}{$\mathbf{F}(\boldsymbol{\Phi}_{l}) = 2 \rm{Re}\bigg\{\frac{\partial \boldsymbol{\mu}^{\rm{H}}_{y_{l}}}{\partial \boldsymbol{\Phi}_{l}} \boldsymbol{\Sigma}_{\mathbf{y}_{l}}^{-1} \frac{\partial \boldsymbol{\mu}_{\mathbf{y}_{l}}}{\partial \boldsymbol{\Phi}_{l}}\bigg\} = \frac{2}{\sigma_{\mathbf{y}_{l}}^2} \begin{bmatrix}
% F_{\boldsymbol{\eta}_{l}\boldsymbol{\eta}_{l}} & F_{\boldsymbol{\eta}_{l} \theta_{l}} & F_{\boldsymbol{\eta}_{l} \bar{\alpha}_{l}} \\
% F_{\boldsymbol{\eta}_{l} \theta_{l}}^{\text{T}} & F_{\theta_{l}\theta_{l}} & F_{\bar{\alpha}_{l}\theta_{l}}  \\
% F_{\boldsymbol{\eta}_{l} \bar{\alpha}_{l}} ^{\text{T}} & F_{\bar{\alpha}_{l}\theta_{l}}^{\text{T}} & F_{\bar{\alpha}_{l}\bar{\alpha}_{l}}
% \end{bmatrix}.$}
% \end{align}
A derivation of the CRB relation can be found in \cite{kay1993fundamentals}. We are mainly interested in the CRB for the delay-Doppler and angle parameters.
% \textcolor{red}{Write the general form of the CRB. From Fisher... Then Write the simplification with delay-Doppler angle separation. Then write the CRB for delay-Doppler-angle is block diagonal. Then write the CRB eta as we had before. Continue. ALPHA'nın tanımını da yazmam lazım. İlerideki paragrafları silme}
% \subsection{Performance Bound for Parameter Estimation}
We group the temporal parameters delay and Doppler shift and denote them with $\boldsymbol{\eta}_{l} = [\tau_{l}, \ \nu_{l}]^{\top} $ for each AP $l$. We assume that  $\bar{\alpha}_{l}$ is an unknown deterministic constant.
 As the signal and noise models are space-time separable, the angle and delay-Doppler CRBs are decoupled, resulting in the block-diagonal CRB relation \cite{dogandzic2001cramer}
 \begin{align}    \text{CRB}_{\boldsymbol{\eta}_{l}\theta_{l}}=\begin{bmatrix}
         \text{CRB}_{\boldsymbol{\eta}_{l}} & \mathbf{0} \\ \mathbf{0}&\text{CRB}_{\theta_{l}}
     \end{bmatrix}.
 \end{align}
 The CRB for the delay-Doppler shift and angle estimation for our model can be obtained as
\begin{align}
    &\text{CRB}_{\boldsymbol{\eta}_{l}} = 
    \begin{bmatrix}
        \text{CRB}_{\tau_{l} \tau_{l}}  & \text{CRB}_{\tau_{l} \nu_{l}} \\
        \text{CRB}_{\tau_{l} \nu_{l}} & \text{CRB}_{\nu_{l} \nu_{l}}
    \end{bmatrix}  \notag  \\ \label{eq:CRB delay-Doppler shift} & \resizebox{\linewidth}{!}{$=\left( \frac{2 |\bar{\alpha}_l|^2 \Vert \mathbf{a}(\theta_{l})\Vert^2}{\sigma_{n}^2} \operatorname{Re} \left\{ \frac{\partial \breve{\boldsymbol{\varsigma}}^{
    }(\boldsymbol{\eta}_l)}{\partial \boldsymbol{\eta}_l}\left( \mathbf{I}_{N} - \frac{\breve{\boldsymbol{\varsigma}}(\boldsymbol{\eta}_l) \breve{\boldsymbol{\varsigma}}^{H}(\boldsymbol{\eta}_l) }{ \Vert\breve{\boldsymbol{\varsigma}}(\boldsymbol{\eta}_l)\Vert^2} \right)\frac{\partial \breve{\boldsymbol{\varsigma}}(\boldsymbol{\eta}_l)}{\partial \boldsymbol{\eta}_l} \right\} \right)^{-1},$}\\
    &\text{CRB}_{\theta_{l}} = \notag\\
    &\resizebox{\linewidth}{!}{$\left( \frac{2 |\bar{\alpha}_l|^2 \Vert \breve{\boldsymbol{\varsigma}}(\boldsymbol{\eta}_l)\Vert^2}{\sigma_{n}^2} \operatorname{Re} \left\{ \frac{\partial \mathbf{a}^{H}(\theta_{l})}{\partial \theta_{l}}\left( \mathbf{I}_{N_c N_s} - \frac{\mathbf{a}(\theta_{l}) \mathbf{a}^{H}(\theta_{l}) }{ \Vert\mathbf{a}(\theta_{l})\Vert^2} \right)\frac{\partial \mathbf{a}(\theta_{l})}{\partial \theta_{l}} \right\} \right)^{-1},$}
\end{align}
% Steps relating \eqref{eq:Fisher information matrix derivation} and \eqref{eq:CRB delay-Doppler shift} are given in Appendix \ref{appendix:Delay-Doppler shift FIM Derivation}.
where $\breve{\boldsymbol{\varsigma}}(\tau_l, \nu_l) = [\breve{\boldsymbol{\varsigma}}_{1}^{\top}(\tau_l, \nu_l), \dots, \breve{\boldsymbol{\varsigma}}_{N_s}^{\top}(\tau_l, \nu_l)]^{\top} \in \mathbb{C}^{N_c N_s \times 1}$, for $\breve{\boldsymbol{\varsigma}}_{b}(\tau_l, \nu_l)=[\breve{\boldsymbol{\varsigma}}_{b}[0](\tau_l, \nu_l), \dots, \breve{\boldsymbol{\varsigma}}_{b}[N_{\rm{c}}-1](\tau_l, \nu_l)]^{\top}\in \mathbb{C}^{N_c \times 1}$. The range and velocity CRBs are found by the transformation of parameters property of CRB \cite{kay1993fundamentals}. 
% The property can be stated as $\rm{CRB}_{\phi_{i'}\phi_{j'}}=\big(\frac{\partial^{2} g}{\partial \phi_{i}\partial \phi_{j}}\big) \rm{CRB}_{\phi_{i}\phi_{j}}$ for parameters related with $(\phi_{i'},\phi_{j'})=g(\phi_{i},\phi_{j})$. We relate the range and radial velocity to delay and Doppler shift as $\frac{ R_{\rm{R}} +  R_{l}}{c}= \tau_{l}$ and $\frac{2v_{r,l} f_{\rm{c}}}{c}=\nu_{l}$, where $f_{\rm{c}}$ is the carrier frequency. Therefore, we have $\mathrm{CRB}_{R_{l}R_{l}}=c^2 \text{CRB}_{\tau_{l} \tau_{l}} $
% , $\mathrm{CRB}_{v_{r,l}v_{r,l}}=(\frac{c}{2f_{\rm{c}}})^2\text{CRB}_{\nu_{l} \nu_{l}}$ and $\mathrm{CRB}_{R_{l}v_{l}}=(\frac{c^2}{2f_{\rm{c}}}) \text{CRB}_{\tau_{l} \nu_{l}} $. 
We form the following CRB matrix
\begin{align}
\text{CRB}_{\boldsymbol{\tilde{\eta}}_{l}\theta_{l}}= 
% \begin{bmatrix}
% \mathrm{CRB}_{R_{l}R_{l}} & \mathrm{CRB}_{R_{l} v_{r,l}}  \\
% \mathrm{CRB}_{R_{l} v_{r,l}} & \mathrm{CRB}_{v_{r,l}v_{r,l}} 
% \end{bmatrix}=
\mathbf{A}\text{CRB}_{\boldsymbol{\eta}_{l}\theta_{l}} \mathbf{A}^{\top} ,
% \quad \mathbf{A} &= \begin{bmatrix} c & 0 & 0 \\ 0 &  c/2f_{c} & 0  \\ 0 & 0 & 1 \end{bmatrix}
\end{align}
where $\mathbf{A}=\mathrm{diag}(c,c/2f_{\rm{c}},1)$, where $c$ is the speed of light, $f_{c}$ is the carrier frequency and $\mathrm{diag}(\cdot)$ denotes the diagonal matrix.
The CRB of all the parameters at all receive APs can be lower bounded with $\text{CRB}_{\boldsymbol{\tilde{\eta}}\theta}=\rm{blkdiag}(\text{CRB}_{\boldsymbol{\tilde{\eta}}_{1}\theta_{1}},\ldots,\text{CRB}_{\boldsymbol{\tilde{\eta}}_{L}\theta_{L}})$.

\section{User Tracking}

% In this section, we introduce a Kalman filtering framework designed for beam tracking. We adopt the constant velocity motion model along the horizontal axis. As APs estimate the radial distance and velocity of the users, there is a coordinate change in the observation equation which poses nonlinearities.
% This prevents the direct use of the linear Kalman filter. Consequently, we employ an extended Kalman filter (EKF) approach, which utilizes local linearization to handle the nonlinearities.
Next, we will present an extended Kalman filtering algorithm for tracking and predicting the location of the user utilizing the estimated radial distances and velocities.

\subsection{Extended Kalman Filter} \label{sec: EKF}
Consider the scenario illustrated in Fig. \ref{fig:APs and car system}, where the APs are uniformly spaced along a horizontal line, while a vehicle travels in the positive direction at a constant speed. We define the state vector of the user as
\begin{align}
    \mathbf{x}_{\kappa} = [p_{x}[\kappa] \ v_{x}[\kappa]]^{\top},
\end{align} 
where $p_{x}[\kappa]$ and  $v_{x}[\kappa]$ are the horizontal position and velocity of the user and $\kappa$ is the time index for the filter. Each epoch $\kappa$ corresponds to $N_{s}$ OFDM symbols, and each OFDM symbol contains $N_{c}$ time samples. Several coherence blocks collectively form the coherent processing interval (CPI), which we use as the epochs.

% Furthermore, as described in Section~\ref{subsec: Sensing Transmitted}, a subset of the APs is designated as the Rx APs.  We use $\{\ell_1, \ldots, \ell_{L_\kappa}\}$ to denote the indicies of these Rx APs at epoch $\kappa$. We only use the measurements obtained from them and define the measurement vector as 
Measurement vector is formed with the measurements from the selected APs
% \textcolor{red}{yanlış set düzgün olmalı}
% \begin{align}
%     \mathbf{z}_{\kappa} = [\hat{R}_{1}[\kappa],\hat{v}_{r,1}[\kappa],\hat{\theta}_{1}[\kappa],\ldots, \hat{R}_{L_{\kappa}}[\kappa],\hat{v}_{r, L_{\kappa}}[\kappa],\hat{\theta}_{L_{\kappa}}[\kappa]]^{\top},   
% \end{align}
% \begin{equation}
% \mathbf{z}_{\kappa} = \begin{bmatrix}
% \hat{R}_{\ell_1}[\kappa] , \hat{v}_{r,\ell_1}[\kappa] , \hat{\theta}_{\ell_1}[\kappa] , \cdots ,
% \hat{R}_{\ell_{L_\kappa}}[\kappa] , \hat{v}_{r,\ell_{L_\kappa}}[\kappa] , \hat{\theta}_{\ell_{L_\kappa}}[\kappa]
% \end{bmatrix}^{\top}\mkern-12mu,
% \end{equation}
% where $\hat{R}_{l}[\kappa]$, $\hat{v}_{r,l}[\kappa]$ and $\hat{\theta}_{l}[\kappa]$ are the range, velocity and angle estimates at Rx AP $l$.
Assuming that the user motion can be modeled by a constant velocity model, the state-space model that describes the system behavior is given by 
\begin{align}\label{eq:state space} \mathbf{x}_{\kappa+1} &= \mathbf{F}\mathbf{x}_{\kappa} + \mathbf{w}_{\kappa}, \quad \mathbf{w}_{\kappa} \stackrel {\textrm{i.i.d.}}\sim \mathcal{N}(\mathbf{0},\mathbf{Q}), \\ \mathbf{z}_{\kappa} &= \mathbf{h}(\mathbf{x}_{\kappa}) + \mathbf{v}_{\kappa}, \quad \mathbf{v}_{\kappa} \stackrel {\textrm{i.i.d.}}\sim \mathcal{N}(\mathbf{0},\mathbf{R}_{\kappa}). \end{align}
% State vector is
% \begin{align} \mathbf{x}_{\kappa} = \begin{bmatrix} p_{x}[\kappa] \\ v_{x}[\kappa]  \end{bmatrix} \end{align}
% Measurement vector is
% \begin{align} \mathbf{z}_{\kappa} = \begin{bmatrix} R[\kappa] \\ v_{r}[\kappa]  \end{bmatrix} \end{align}
The state transition matrix $\mathbf{F}$ and the process noise covariance $\mathbf{Q}$ for the constant velocity motion are defined as \cite{1261132}
\begin{align}
    % Process model (constant velocity)
\mathbf{F} &= \begin{bmatrix} 1 & \Delta_{T} \\ 0 & 1 \end{bmatrix}, \quad \mathbf{Q} = \begin{bmatrix}
\frac{\Delta_{T}^{4}}{4}\sigma_{q}^{2} & \frac{\Delta_{T}^{3}}{2}\sigma_{q}^{2} \\
\frac{\Delta_{T}^{3}}{2}\sigma_{q}^{2} & \Delta_{T}^{2}\sigma_{q}^{2}
\end{bmatrix},
\end{align}
where $\Delta_{T}$ is the time step between epochs and $\sigma_{q}^{2}$ is the variance of the process noise that represents the uncertainty in the target's acceleration. 
% \begin{align}
% Q = \begin{bmatrix}
% \frac{\Delta_{T}^{4}}{4}\sigma_{q}^{2} & \frac{\Delta_{T}^{3}}{2}\sigma_{q}^{2} \\
% \frac{\Delta_{T}^{3}}{2}\sigma_{q}^{2} & \Delta_{T}^{2}\sigma_{q}^{2}
% \end{bmatrix},
% \end{align}
The measurement relation $\mathbf{h}(\mathbf{x}_\kappa)$ only contains the contributions from the selected APs where each entry is
% \begin{subequations}
% % \begin{align}
% %     \mathbf{h}(\mathbf{x}_\kappa)=\Big[[\mathbf{h}(\mathbf{x}_\kappa)]_{1}^{\top},\ldots,[\mathbf{h}(\mathbf{x}_\kappa)]_{L_{\kappa}}^{\top}\Big],
% % \end{align}
% \begin{equation}
% \mathbf{h}(\mathbf{x}_\kappa) = \begin{bmatrix}
% [\mathbf{h}(\mathbf{x}_\kappa)]_{\ell_1}^{\top},  
% \ldots, 
% [\mathbf{h}(\mathbf{x}_\kappa)]_{\ell_{L_\kappa}}^{\top}
% \end{bmatrix},
% \end{equation}
\begin{align}
\resizebox{\linewidth}{!}{$[\mathbf{h}(\mathbf{x}_\kappa)]_{l}=\Bigg[\sqrt{((p_x[\kappa] - p_{l})^{2}+ p_{y})^{2}}  \ \frac{(p_x[\kappa] - p_{l}) v_x[\kappa]}{\sqrt{((p_x[\kappa] - p_{l})^{2}+ p_{y})^{2}}}\Bigg],$}
\end{align}
% \end{subequations}
%ALTERNATIVE
% $[\mathbf{h}(\mathbf{x}_\kappa)]_{l}=\Bigg[\sqrt{(p_x[\kappa]-p_l)^2+p_{y}^2} 
% \dfrac{(p_x[\kappa]-p_l)\,v_x[\kappa]}{\sqrt{(p_x[\kappa]-p_l)^2+p_{y}^2}}\Bigg]$
% $\mathbf{h}(\mathbf{x}_\kappa) = 
% \begin{bmatrix}
% \sqrt{(p_x[\kappa]-p_1)^2+p_{y}^2}, & 
% \dfrac{(p_x[\kappa]-p_1)\,v_x[\kappa]}{\sqrt{(p_x[\kappa]-p_1)^2+p_{y}^2}}, & 
% \sqrt{(p_x[\kappa]-p_2)^2+p_{y}^2}, & 
% \dfrac{(p_x[\kappa]-p_2)\,v_x[\kappa]}{\sqrt{(p_x[\kappa]-p_2)^2+p_{y}^2}}, & 
% \ldots, & 
% \sqrt{(p_x[\kappa]-p_L)^2+p_{y}^2}, & 
% \dfrac{(p_x[\kappa]-p_L)\,v_x[\kappa]}{\sqrt{(p_x[\kappa]-p_L)^2+p_{y}^2}}
% \end{bmatrix}^{\top}$
%%%%%%%%%%%%%% Observation equaiton
% \begin{align}
% \mathbf{h}(\mathbf{x}_\kappa)=
% \begin{bmatrix}
% \sqrt{(p_x[\kappa]-p_1)^2+p_{y}^2} \\[1mm]
% \dfrac{(p_x[\kappa]-p_1)\,v_x[\kappa]}{\sqrt{(p_x[\kappa]-p_1)^2+p_{y}^2}} \\[2mm]
% \sqrt{(p_x[\kappa]-p_2)^2+p_{y}^2} \\[1mm]
% \dfrac{(p_x[\kappa]-p_2)\,v_x[\kappa]}{\sqrt{(p_x[\kappa]-p_2)^2+p_{y}^2}} \\[2mm]
% \vdots \\[2mm]
% \sqrt{(p_x[\kappa]-p_{L_{\kappa}})^2+p_{y}^2} \\[1mm]
% \dfrac{(p_x[\kappa]-p_{L_{\kappa}})\,v_x[\kappa]}{\sqrt{(p_x[\kappa]-p_{L_{\kappa}})^2+p_{y}^2}}
% \end{bmatrix},
% \end{align}
where $p_{y}$ is the fixed vertical distance of the target from the APs. 
% The EKF observation matrix is obtained by taking the Jacobian of the observation equation as
% \begin{align}
%     \resizebox{\linewidth}{!}{$\mathbf{H}_\kappa=
% \begin{bmatrix}
% \dfrac{p_x[\kappa]-p_1}{\sqrt{(p_x[\kappa]-p_1)^2+p_{y}^2}} & 0 \\[1mm]
% \dfrac{v_x[\kappa]\,p_{y}^2}{\left((p_x[\kappa]-p_1)^2+p_{y}^2\right)^{3/2}} & \dfrac{p_x[\kappa]-p_1}{\sqrt{(p_x[\kappa]-p_1)^2+p_{y}^2}} \\[2mm]
% \dfrac{p_x[\kappa]-p_2}{\sqrt{(p_x[\kappa]-p_2)^2+p_{y}^2}} & 0 \\[1mm]
% \dfrac{v_x[\kappa]\,p_{y}^2}{\left((p_x[\kappa]-p_2)^2+p_{y}^2\right)^{3/2}} & \dfrac{p_x[\kappa]-p_2}{\sqrt{(p_x[\kappa]-p_2)^2+p_{y}^2}} \\[2mm]
% \vdots & \vdots \\[2mm]
% \dfrac{p_x[\kappa]-p_{L_{\kappa}}}{\sqrt{(p_x[\kappa]-p_{L_{\kappa}})^2+p_{y}^2}} & 0 \\[1mm]
% \dfrac{v_x[\kappa]\,p_{y}^2}{\left((p_x[\kappa]-p_{L_{\kappa}})^2+p_{y}^2\right)^{3/2}} & \dfrac{p_x[\kappa]-p_{L_{\kappa}}}{\sqrt{(p_x[\kappa]-p_{L_{\kappa}})^2+p_{y}^2}}
% \end{bmatrix}.
% $}
% \end{align}
We set the measurement noise covariance matrix based on the CRB relation from Section~\ref{sec:sensingCRB}, i.e., $ \mathbf{R}_{\kappa} = \text{CRB}_{\boldsymbol{\tilde{\eta}}_{\kappa}\theta_{\kappa}} $.
Given the state space model in \eqref{eq:state space}, the state $\mathbf{x}_{\kappa}$ can be recursively estimated using the EKF algorithm in Table \ref{table:EKF equations}. Note that the EKF recursions predict the covariance matrix even without new observations.

\begin{table}[t!]
\centering
\caption{EKF equations used to estimate user state and uncertainty (covariance) of the estimate.}
\label{table:EKF equations}
\begin{tabular}{rl}
\hline
\multicolumn{2}{c}{\textbf{Time update (prediction step)}} \\
\hline
State prediction & $\hat{\mathbf{x}}_{\kappa|\kappa-1} = \mathbf{F}\mathbf{x}_{\kappa-1|\kappa-1}$ \\
Covariance prediction & $\mathbf{P}_{\kappa|\kappa-1} = \mathbf{F}\mathbf{P}_{\kappa-1|\kappa-1}\mathbf{F}^{\top} + \mathbf{Q}$ \\
\hline
\multicolumn{2}{c}{\textbf{Measurement update}$^{*}$} \\
\hline
EKF observation matrix & $\mathbf{H}_{\kappa} = \nabla \mathbf{h}(\mathbf{x})\bigg\rvert_{\mathbf{x}=\mathbf{x}_{\kappa|\kappa-1}}$ \\
Innovation & $\bar{\mathbf{z}}_{\kappa} = \mathbf{z}_{\kappa} - h(\hat{\mathbf{x}}_{\kappa|\kappa-1})$ \\
Innovation covariance & $\mathbf{S}_{\kappa} = \mathbf{H}_{\kappa}\mathbf{P}_{\kappa|\kappa-1}\mathbf{H}_{\kappa}^{\top} + \mathbf{R}$ \\
Kalman gain & $\mathbf{K}_{\kappa} = \mathbf{P}_{\kappa|\kappa-1}\mathbf{H}_{\kappa}^{\top}\mathbf{S}_{\kappa}^{-1}$ \\
State update & $\hat{\mathbf{x}}_{\kappa|\kappa} = \hat{\mathbf{x}}_{\kappa|\kappa-1} + \mathbf{K}_{\kappa}\bar{\mathbf{z}}_{\kappa}$ \\
Covariance update & $\mathbf{P}_{\kappa|\kappa} = (\mathbf{I} - \mathbf{K}_{\kappa}\mathbf{H}_{\kappa})\mathbf{P}_{\kappa|\kappa-1}$ \\
\hline
\end{tabular}\\[1mm]
{\footnotesize {*} Measurement update is done only when in the sensing state.}
\end{table}

We relate the position of the target to the angle of it from the origin by the following relation
\begin{align}
    g(\hat{p}_x)=\arctan\left(\frac{\hat{p}_x}{p_{y}}\right).
\end{align}
From the estimated user position, the estimate of the direction (angle) to the user can be calculated as
\begin{align}
    \hat{\theta}^{ \text{\tiny EKF}}[\kappa] = g(\hat{p}_x[\kappa|\kappa^{*}]).
\end{align}
% Further, the uncertainty in the estimated angle can be approximated as
The associated estimation error variance can be approximated as
\begin{align}
    \mathrm{var}(\tilde{\theta}^{\text{\tiny EKF}}[\kappa|\kappa^{*}]) = [\mathbf{P}_{\kappa|\kappa^{*}}]_{1,1}\left(\frac{\partial g}{\partial p_{x}}\Biggr\rvert_{p_{x}=\hat{p}_{x}[\kappa|\kappa^{*}]}\right)^{2},
    % \frac{\partial g}{\partial p_{x}}\Biggr\rvert_{p_{x}=\hat{p}_{x}}
\end{align}

% where $\mathbf{g}=[\arctan\left(g(\hat{p}_x[\kappa|\kappa^{*}])\right), 0]^{\top}$ and $\boldsymbol{\omega}_{\kappa}$ is the AP selection vector explained in the next section.

\section{Sensing Management}\label{sec: sensing management}
Next, we propose methods to choose when to do sensing and which APs will then be used as Rx APs. These actions are made based on the variance of the predicted direction (angle) estimation error to the user, calculated using the EKF.
% The outline of the process can be seen in Fig. \ref{fig:systemmodel}.

To implement the Rx AP selection we introduce the vector $\boldsymbol{\omega}_{\kappa}=[\omega_{1,\kappa},\ldots,\omega_{L_{\mathrm{T}},\kappa}]^{\top}\in \mathbb{Z}^{L_{\mathrm{T}}\times1}_{2}$, where $\omega_{l,\kappa}\in\{0,1\} \ \text{for} \ l=1,\ldots, L_{\mathrm{T}}$. We determine that AP $l$ is an Rx AP if $\omega_{l,\kappa}=1$ and unused otherwise. The number of Rx APs is $L_{\kappa}$.
% The selected APs form the Rx APs set $ \mathcal{R}_{\kappa}$. The cardinality of this set is $ \vert\mathcal{R}_{\kappa}\vert=L_{\kappa}$. 

% \subsection{Sensing Time and Receive AP Selection} \label{subsec: Sensing Frequency Selection}
Herein, we consider the times that we need to transmit sensing signals and do the tracking to keep a certain performance on the angle estimate. We would like the estimated angles to be in the half-power beamwidth to keep an effective angular resolution. Therefore, radar signals are transmitted based on a comparison between the predicted variance of the angle estimation error and a threshold derived from the half-power beamwidth, as detailed in the appendix. We choose the action for the next epoch as
\begin{align}
    \rm{Action}[\kappa+1] = \begin{cases*}
  \rm{Sensing}, & if  $ \rm{var}(\tilde{\theta}^{\text{\tiny EKF}}[\kappa+1|\kappa^{*}]) > \gamma$,\\
  \rm{No \ Sensing},              & if $\rm{var}(\tilde{\theta}^{\text{\tiny EKF}}[\kappa+1|\kappa^{*}]) < \gamma$,
  \end{cases*}
\end{align}
where $\kappa^{*}$ is the last time sensing was done. The predicted estimation error variance is 
\begin{align}
    \rm{var}(\tilde{\theta}^{\text{\tiny EKF}}[\kappa+1|\kappa^{*}]) &= [\mathbf{P}_{\kappa+1|\kappa^{*}}(\boldsymbol{\omega}_{\kappa})]_{1,1}\nonumber\\ &\cdot\left(\frac{\partial g}{\partial p_{x}}\Bigr\rvert_{p_{x}=\hat{p}_{x}[\kappa+1|\kappa^{*}]}\right)^{2}.
\end{align}
It is obtained by using the last Rx AP set.
% \textcolor{red}{Explain how this is done in the next step, how is the variance found first then time (AP selection is before).}
% \subsection{}\label{sec:Receive AP Selection}
We then consider the receive AP selection problem to maximize the tracking and communication performance. To achieve this, we consider minimizing the predicted angle estimation variance as a function of the AP selection matrix at the next epoch $\boldsymbol{\omega}_{\kappa+1}$, which can be written as
\begin{align} \label{eq: Angle estimation error variance with AP selection}
\boldsymbol{\omega}_{\kappa+1}
% &= \argmin_{\breve{\boldsymbol{\omega}}\in\mathbb{Z}^{\rm{L_{Tot}\times 1}}_{2}}\rm{var}(\tilde{\theta}^{\text{\tiny EKF}}[\kappa+1|\kappa^{*}]) \notag\\
&= \argmin_{\breve{\boldsymbol{\omega}}\in\rm{\mathbb{Z}^{\rm{L_{Tot}\times 1}}_{2}}} \rm{var}(\tilde{\theta}^{\text{\tiny EKF}}[\kappa+1|\kappa^{*}])
% [\mathbf{P}_{\kappa+1|\kappa^{*}}(\breve{\boldsymbol{\omega}})]_{1,1}\left(\frac{\partial g}{\partial p_{x}}\Biggr\rvert_{p_{x}=\hat{p}_{x}[\kappa+1|\kappa^{*}]}\right)^{2}.
\end{align}
% The variance of the estimation error is obtained from the predicted MSE matrix as 
% \begin{align} \label{eq: Angle estimation error variance}
%     \rm{var}(\tilde{\theta}^{\tiny \mathrm{EKF}}[\kappa+1|\kappa]) = \nabla\mathbf{g} [\mathbf{P}_{\kappa+1|\kappa}]_{1,1}\nabla\mathbf{g}^{\top}.
% \end{align}
% \begin{align} \label{eq: Angle estimation error variance}
%     \rm{var}(\tilde{\theta}^{\tiny \mathrm{EKF}}[\kappa+1|\kappa+1]) = \nabla\mathbf{g} [\mathbf{P}_{\kappa+1|\kappa+1}(\boldsymbol{\omega}_{\kappa})]_{1,1}\nabla\mathbf{g}^{\top}.
% \end{align}
To choose the selection vectors $\boldsymbol{\omega}_{\kappa+1}$, we look at all $2^{{L}_{\rm{T}}}$ possible choices and pick the one that minimizes the predicted angle estimation error variance.  
% We note that the condition $\vert \mathcal{R}_{\kappa+1}\vert + \vert \mathcal{T}_{\kappa+1}\vert = {L}_{\rm{T}}$ needs to be satisfied.

% \begin{subequations}
% \begin{alignat}{2}
% &\!\min_{\boldsymbol{\omega}_{\kappa+1}}        &\qquad& \mathcal{{L}}(\boldsymbol{\omega}_{\kappa+1})\label{eq:optProb}\\
% &\text{subject to} &      & \vert \mathcal{R}_{\kappa+1}\vert + \vert \mathcal{T}_{\kappa+1}\vert = {L}_{\rm{T}},\label{eq:constraint1}\\
% &                  &      & \omega_{l_{t}}\in\{0,1\} \ for \ l=1,\ldots,{L}_{\rm{T}} .\label{eq:constraint2}
% \end{alignat}
% \end{subequations}

\section{Simulation Setting and Results}
We evaluate the performance of the proposed predictive beamforming in a cell-free massive MIMO system similar to the one shown in Fig.~\ref{fig:APs and car system}. In our scenario, we assume that the APs are located equidistantly on a horizontal line along the road and a vehicle moves in the positive direction with a constant velocity. The setup consists of $L_{\rm{T}}$ = 4 APs with $N$ = 4 antennas. Unless otherwise noted, the system parameters are as given in Table~\ref{tab:sim_params}.
% align with the framework in \cite{akçalı2025predictivebeamformingdistributedmimo}. 
% A scenario and system setup as illustrated in Fig. \ref{fig:APs and car system} is considered. 
% The APs are located equidistantly on a $\SI[per-mode=fraction]{500}{\meter}$ line. The vehicle initiates at the position $p_{x}[0]=\SI[per-mode=fraction]{0}{\meter}$ and travels in the positive x-direction at a constant speed $v_{x}[0]=25 \, m/s$. 
The initial uncertainty in the position and velocity of the user is characterized by the covariance matrix $\mathbf{P}_{0|0}=\mathrm{diag}(100,1)$. The sensing time selection threshold is $\gamma = 3^{\circ}$. We assume that the RCS fluctuates according to the Swerling I model with a mean value $ 5\ \textrm{m}^ {2}$.

% Then for your table
\begin{table}[t!]
\centering
\caption{SIMULATION PARAMETERS}
\label{tab:sim_params}
\begin{tblr}{|l|c|}
\hline
Parameter & Value \\
\hline
Number of AP antennas $(N)$ & 4 \\
\hline
Number of APs ($L$) &  4 \\
\hline
Time difference  & $\triangle_T=0.01$ \\
\hline
Acceleration noise parameter  & $ \sigma_q=0.1$ \\
\hline
Position of the $l$th AP $(x,y)$ & $((500/L)l,0)$\\
\hline
Vehicle start position $(x,y)$ & $(0,40)$ \\
\hline
Carrier frequency $(f_c)$ & 30 GHz \\
\hline
Number of epochs & 200 \\
\hline
Velocity $(v)$ & 25 m/s \\
\hline
Noise variance $(\sigma_n^2)$ & $-75$ dBm \\
\hline
Transmit power $(\rho_{d})$ & 1 W \\
\hline
\end{tblr}
\end{table}

\subsection{Instantaneous Downlink Rates} \label{subsec: Comparison of SNR - and forming the precoder}
Upon transitioning to the ON state, the communication signal is transmitted in the downlink using a maximum ratio (MR) precoder $\mathbf{w}_{\kappa}=\sqrt{\rho_{d}/N}\mathbf{a}(\hat{\theta}^{\text{\tiny EKF}}[\kappa])$. The angle $\hat{\theta}^{\text{\tiny EKF}}[\kappa]$ is the predicted angle using all the sensing observations up to the last sensing state $\kappa^{*}$. 
% On the other hand, $\hat{\theta}^{\mathrm{FFT}}[\kappa]$ corresponds to the angle estimate obtained with the FFT method during the previous sensing state $\kappa^{*}$. 

The instantaneous downlink rate, assuming perfect CSI at the receiver during the decoding, is computed as
\begin{equation}
R^{\mathrm{EKF}}_{\kappa} = \rm{log}_{2} (1+\mathrm{SNR}_{\kappa}^{\mathrm{EKF}}),
\end{equation}
where $\mathrm{SNR}^{\mathrm{EKF}}_{\kappa}$ is the SNR in \eqref{eq:SNRexpression} obtained at epoch $\kappa$ using $\mathbf{w}_{\kappa}$. The average capacity can be obtained by averaging over the channel realizations across several epochs.
% \textcolor{red}{However, we have the information on angle estimation error variance and the estimates between all APs and the OFF users. Therefore, we can form other precoders.} 

\subsection{Simulation Results}

\begin{figure}[t!]
\centering
  \includegraphics[width=\linewidth]{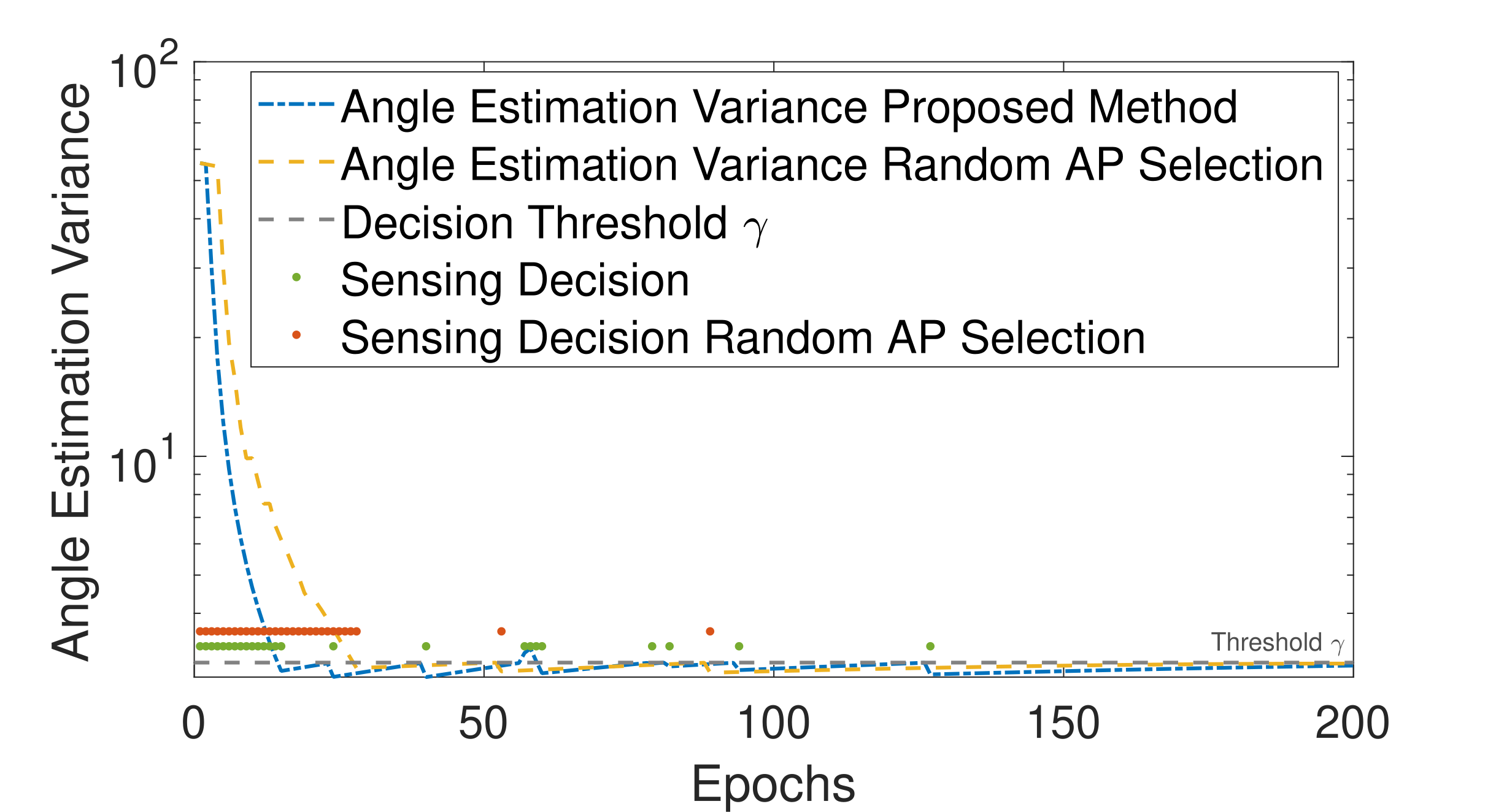}\vspace{-3mm}
    \caption{The temporal behavior of the predicted angle estimation variance and sensing decisions with optimal and random Rx AP selection.}\vspace{-3mm}
    \label{fig:angle_cov}
\end{figure}

\begin{figure}[t!]
\centering
  \includegraphics[width=\linewidth]{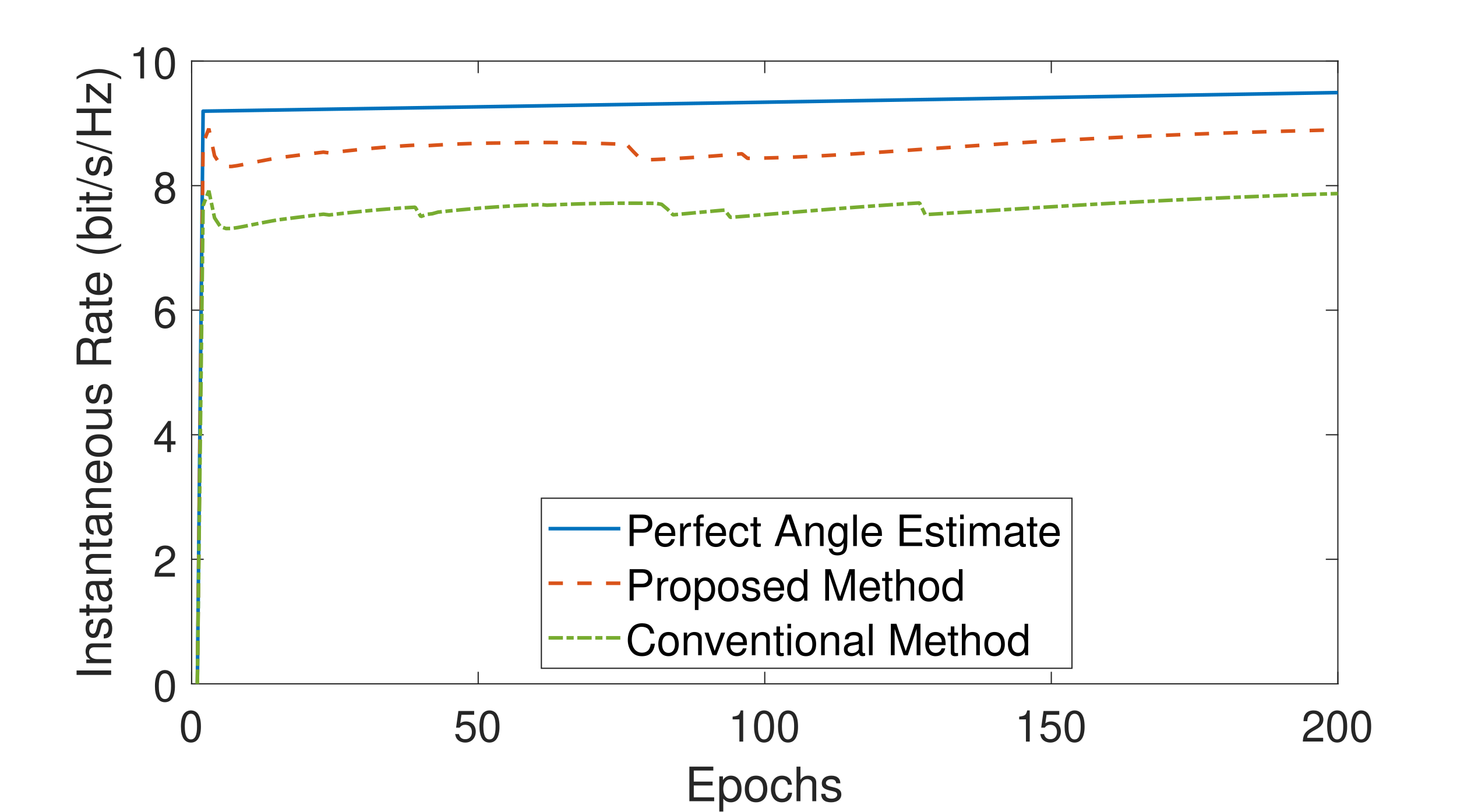}\vspace{-2mm}
    \caption{The temporal behavior of the instantaneous rate with and without sensing management and the perfect angle estimates with $L=4$, $N=4$.}
    \label{fig:se_time}\vspace{-3mm}
\end{figure}

% \textcolor{red}{Fig.~\ref{fig:tracking} depicts the tracking accuracy for angle and distance parameters. } 
Fig.~\ref{fig:angle_cov} illustrates the temporal behavior of the predicted angle estimation variance with the proposed and random Rx AP selection. For the latter, APs are selected randomly from the set of all possible APs. Moreover, Fig.~\ref{fig:se_time} shows the temporal evolution of the instantaneous rate the system can support if there is a communication request at a particular epoch. We consider the downlink rates achieved with the proposed method, the conventional case, and the perfect angle knowledge at APs. For the proposed method, we use the angle obtained by the EKF algorithm. For the conventional case,  as the proposed sensing management isn't used, a power allocation between sensing and communication is considered. Half of the transmit power $\rho_{d}$ is used for communication, and the other half is reserved for the radar signal transmission. 
% We consider both the FFT and the EKF methods for the cell-free massive MIMO setup. 
% For FFT based angle estimation, a grid spanning from $-90^{\circ}$ to $90^{\circ}$ in $1^{\circ}$ increments is considered.
% We also note that only a portion of these APs are used in sensing which is selected predictively as explained in section \ref{sec:Receive AP Selection}.

% %%%%%%%%%%%%%%% ADDITIONAL FIGURE
% \textcolor{red}{In Fig.~\ref{fig:tracking}, one can observe that the tracked values of distance and angle exhibit strong alignment with their true counterparts. This accuracy shows itself in the SE relations.}

% \begin{figure}[t!]
% \centering
%   \includegraphics[width=\linewidth]{Track_fig1.eps}\vspace{-6mm}
%     \caption{Tracking of target's position, velocity, and angle.}\vspace{-3mm}
%   \label{fig:tracking}
% \end{figure}

Fig.~\ref{fig:angle_cov} shows that the angle estimation variance drops significantly when there is a sensing decision in the former epoch. We can see that there isn't a necessity for frequently transmitting sensing signals while maintaining highly accurate angle estimates, except in the beginning because the initial uncertainty was assumed to be high. However, we see that there is a more frequent need for sensing without the proposed AP selection algorithm. It takes about twice the time to fall below the threshold with the random AP selection.

Fig.~\ref{fig:se_time} shows that the proposed method outperforms the conventional method with the power allocation we assumed.
The proposed method with the cell-free massive MIMO system maintains a stable communication performance over time. This stems from the broader geographical coverage due to distributed APs. Moreover, the performance of the proposed method is close to the case with the perfect angle estimates. The fluctuations in the rate are caused by the angle estimate diverging from the correct value until it triggers the sensing decision.
% We can see that the EKF method gives a rather constant rate over the epochs.
Fig.~\ref{fig:angle_cov} and Fig.~\ref{fig:se_time} show that channel estimation isn't needed frequently, or reactively when there is data to transmit. Instead, one can track the UE and do sensing when there is an increase in the angle estimation error variance and use the tracked angle for communication when necessary, as proposed.

\section{Conclusion}
We propose an integrated sensing and communication framework for cell-free massive MIMO systems with focus on practical bursty traffic. It is a state-based method whereby user communication requests define the states, and target angles are tracked but sensing signals are only sent when necessary. Predictive beamforming is employed to prevent the necessity for continuous channel estimation, thereby reducing channel estimation overhead. We implement EKF for user tracking. We develop a novel sensing management algorithm, where the sensing time selection and receive AP selection, further improves the overall communication performance by only performing sensing when the estimation variance becomes high and then using the best set of sensing receivers. Numerical evaluations validate the effectiveness of the proposed framework, demonstrating better performance than that of the conventional method, where sensing and communication compete for resources. Moreover, they show that overhead-free communication with predictive beamforming is feasible, and they confirm that continuous sensing is not needed.

\appendix
\section{Appendix}
\label{appendix:The threshold for variance of angle estimate}
The angle estimates from the EKF algorithm can be written as
\begin{equation}
\hat{\theta} \sim \mathcal{N}(\theta_0, \operatorname{var}(\tilde{\theta})),
\end{equation}
where $\theta_0$ is the mean value of the estimate and since $\hat{\theta} - \theta_0$ is normally distributed with zero mean and variance $\operatorname{var}(\tilde{\theta})$.
We wish to ensure that
$
\Pr\Bigl(|\hat{\theta} - \theta_0| > \theta_{\rm{HPBW}}\Bigr) < \epsilon$ for the half-power beamwidth $\theta_{\rm{HPBW}}$ and for any chosen $\epsilon$. We can define the standardized variable
$
\frac{\hat{\theta} - \theta_0}{\sqrt{\operatorname{var}(\tilde{\theta})}} \sim \mathcal{N}(0,1)
$ and write
\begin{equation}
\Pr\Bigl(|\hat{\theta} - \theta_0| > \theta_{\rm{HPBW}}\Bigr)
=2\left[1-\Phi\left(\frac{\theta_{\rm{HPBW}}}{\sqrt{\operatorname{var}(\tilde{\theta})}}\right)\right],
% \Pr\Bigl(|Z| > \frac{\theta_{\rm{HPBW}}}{\sqrt{\operatorname{var}(\tilde{\theta})}}\Bigr)
% =
\end{equation}
where $\Phi(\cdot)$ denotes the cumulative distribution function of the standard normal distribution. Rearranging and solving for $\operatorname{var}(\tilde{\theta})$, the threshold on the variance is then given by
\begin{equation}
\operatorname{var}(\tilde{\theta}) < \left(\frac{\theta_{\rm{HPBW}}}{\Phi^{-1}\Bigl(1-\frac{\epsilon}{2}\Bigr)}\right)^2.
\end{equation}
We denote the right side by $\gamma$.

\bibliographystyle{IEEEtran}
\bibliography{IEEEabrv,isac}

\end{document}